\documentclass[journal]{IEEEtran}
\usepackage{epsfig,latexsym}
\usepackage{float}
\usepackage{indentfirst}
\usepackage{amsmath}
\usepackage{amssymb}
\usepackage{times}
\usepackage{subfigure}
\usepackage{psfrag}
\usepackage{cite}
\usepackage{color}
\usepackage{setspace}
\usepackage{mathrsfs}
\usepackage{stfloats}
\usepackage{graphicx}
\usepackage{algorithm}
\usepackage{algorithmic}
\usepackage[justification=centering]{caption}
\usepackage{multirow}
\usepackage{tabularx}
\usepackage{url}
\usepackage{booktabs}
\usepackage{makecell}

\newtheorem{remark}{Remark}
\newtheorem{theorem}{Theorem}

\newtheorem{lemma}{Lemma}

\newtheorem{corollary}{Corollary}

\usepackage{CJK}

\begin{document}

\title{\huge{Reconfigurable Intelligent Surface aided Integrated-Navigation-and-Communication in Urban Canyons: A Satellite Selection Approach}}

\author{Tianwei Hou, \emph{Member, IEEE}, Da Guan, Xin Sun, Anna Li, \emph{Member, IEEE}, Wenqiang Yi, \emph{Member, IEEE}, Yuanwei Liu, \emph{Fellow, IEEE},  Arumugam Nallanathan, \emph{Fellow, IEEE}

\thanks{This work was supported in part by the Fundamental Research Funds for the Central Universities under Grant 2023JBZY012 and KWJBMC24002536, in part by the National Natural Science Foundation for Young Scientists of China under Grant 62201028, in part by Young Elite Scientists Sponsorship Program by CAST under Grant 2022QNRC001, in part by the Beijing Natural Science Foundation L232041, and in part by the Marie Skłodowska-Curie Fellowship under Grants 101106428 and 101154499. (Corresponding author: Anna Li.)}
\thanks{T. Hou is with the School of Electronic and Information Engineering, Beijing Jiaotong University, Beijing 100044, China, and also with the School of Electronic Engineering and Computer Science, Queen Mary University of London, London E1 4NS, U.K., and also with the Institute for Digital Communications, Friedrich-Alexander Universität Erlangen-Nürnberg (FAU), 91054 Erlangen, Germany (email: twhou@bjtu.edu.cn).}
\thanks{Da Guan and Xin Sun are with the School of Electronic and Information Engineering, Beijing Jiaotong University, Beijing 100044, China(e-mail:23111014@bjtu.edu.cn; xsun@bjtu.edu.cn).}
\thanks{Anna Li is with the School of Computing and Communications, Lancaster University, Lancaster LA1 4WA, U.K. (e-mail: a.li16@lancaster.ac.uk).}
\thanks{Wenqiang Yi is with the School of Computer Science and Electronic Engineering, University of Essex, Colchester CO4 3SQ, U.K. (e-mail: w.yi@essex.ac.uk).}
\thanks{Yuanwei Liu is with the Department of Electrical and Electronic Engineering, The University of Hong Kong, Hong Kong (e-mail: yuanwei@hku.hk).}
\thanks{Arumugam Nallanathan is with the School of Electronic Engineering and Computer Science, Queen Mary University of London, London E1 4NS, U.K. (e-mail: a.nallanathan@qmul.ac.uk).}
}


\maketitle

\begin{abstract}
This study investigates the application of a simultaneous transmitting and reflecting reconfigurable intelligent surface (STAR-RIS)-aided medium-Earth-orbit (MEO) satellite network for providing both global positioning services and communication services in the urban canyons, where the direct satellite-user links are obstructed. Superposition coding (SC) and successive interference cancellation (SIC) techniques are utilized for the integrated navigation and communication (INAC) networks, and the composed navigation and communication signals are reflected or transmitted to ground users or indoor users located in urban canyons. To meet diverse application needs, navigation-oriented (NO)-INAC and communication-oriented (CO)-INAC have been developed, each tailored according to distinct power allocation factors. We then proposed two algorithms, namely navigation-prioritized-algorithm (NPA) and communication-prioritized-algorithm (CPA), to improve the navigation or communication performance by selecting the satellite with the optimized position dilution of precision (PDoP) or with the best channel gain. The effectiveness of the proposed STAR-RIS-aided INAC network is quantified by analyzing the positioning error for navigation services and by evaluating communication performance through achievable ergodic rate metrics.
Our satellite selection approach indicates that: the positioning services at the urban canyon users can be completed with the aid of STAR-RIS. 2) Additionally, it is observed that while a single STAR-RIS array can extend the navigational link, it fails to serve users in indoor scenarios, highlighting a limitation in the current system design.
\end{abstract}


\section{Introduction}

The global navigation satellite system (GNSS), renowned for providing robust all-weather positioning and precise time services, plays a critical role in a myriad of contemporary applications~\cite{t000,t0,t00}. GNSS is integral to various consumer devices, providing centimeter-level positioning accuracies through GNSS carrier phase measurements~\cite{t30}. By 2020, the standards for positioning accuracy have evolved from the 10 cm 2-dimensional (2D) requirement in fifth-generation (5G) systems to an enhanced 1 cm accuracy in 3-dimensional (3D) space for next-generation navigation systems.~\cite{t2}. However, the coverage of GNSS faces challenges when serving areas without direct line-of-sight (LoS) links, such as urban canyons, where the street is flanked by high buildings on both sides creating a canyon-like environment~\cite{t3,t4}.

In GNSS, navigation services are mainly deployed at the medium-earth-orbit (MEO) satellites. Given that medium-earth-orbit (MEO) navigation satellites are situated at altitudes exceeding 20,000 km, the signal power received in open scenarios is approximately -130 dBm~\cite{t31}. In~\cite{t32}, it is indicated that the received signal power is further attenuated by -20 dB in the urban canyons. Additionally, satellite signals reflected from different surfaces in the urban canyon further complicate the evaluation of the pseudo-range via time of flight.
To solve the lack of LoS issues in the urban canyons, existing works have tried from three directions: 1) additional information aided navigation technique; 2) Rayleigh fading channel aided navigation technique; 3) Mega constellation aided navigation technique. For 1), the use of supplementary information, such as 3D maps, also aids navigation when satellite signals are sparse~\cite{line_GNSS_twosatal_with_map}. For 2), in order to provide navigation services in non-LoS (NLoS) cases, the indoor navigation technique based on the Rayleigh fading channel has been proposed, where the indoor distance is evaluated by the multi-path fading with the aid of estimator-correction~\cite{Rayleigh_GNSS_indoor}. The channel statistic can also be predicted in the urban canyons for proving inertial navigation services~\cite{Navigation_predice_urban}.
For 3), In~\cite{LEO_urban_cayon}, since more low-Earth-orbit (LEO) satellites are located at the high pitch angle regimes than that of the MEO satellite in the deep urban canyons, which indicates that the Mega LEO constellations offer better geometric dilution of precision (GDoP) than that of the MEO constellations.
However, these approaches do not create extended LoS links between satellites and users in the urban canyons.

In order to solve the above mentioned issues of the navigation systems, reconfigurable intelligent surface (RIS) emerges as a promising solution. RISs are able to reshape the electromagnetic propagation environment into a controlled format and alter the phase of incoming signals~\cite{t6,t7}. In a word, RISs are able to create new LoS links via reflecting or refracting signals. More specifically, a RIS array consists of numerous passive reflective intelligent surface elements, each containing a reflector layer that can independently induce a phase shift in the electromagnetic wave~\cite{t9,t10,t11}. By doing so, RISs are capable of enhancing the signal power to different wireless applications~\cite{t20,t21,t22,t34,t35}. The RIS-aided 5G networks was proposed, revealing improvements in signal quality and coverage areas~\cite{t27}. An energy efficiency optimization algorithm was proposed in the RIS-aided multi-user networks, where the phase shift of the RIS array was designed~\cite{t19}.
When the RIS array is deployed on the building surfaces, the LoS links between satellite and blocked users in urban canyons can be extended~\cite{Hou_RIS_navigation}. By deploying RIS into the unmanned aerial vehicle, navigation and communications in 3D can be realized~\cite{RIS_UAV_navigation,RIS_UAV_navigation_2}. Traditional RIS implementations have been limited by the requirement that receivers must be located on the same side as the transmitters. In order to release the above constraint of RIS-aided networks, the concept of simultaneous transmitting and reflecting RIS (STAR-RIS) was proposed in~\cite{STAR-RISLiu,STAR-RISXU,STAR-RISMU}. Furthermore, the signal model based on dual-sided STAR-RIS was provided in~\cite{jiaqiSTARRIS_sided}, where the STAR-RIS can achieve full diversity order on both sides of the surface and extend the coverage significantly~\cite{STAR-RISXU}, which has huge development potentials for wireless communication networks. While these innovations have solidified the role of RIS in communication networks, their application in satellite navigation remains unexplored. Utilizing the advanced STAR-RIS technique holds the potential to revolutionize navigation accuracy in urban canyons, marking a significant advancement over current satellite navigation solutions.

Recently, the integrated navigation and communication (INAC) technique is expected to be considered to meet future requirements of satellite networks. Given their relatively lower altitude, LEO satellites are well-suited for communication services but fall short in providing precise navigation due to inadequate Dilution of Precision (DoP). Hence, INAC systems are more favorably deployed on MEO satellites. High-altitude platforms have been proposed for simultaneously delivering cellular communication and navigation services, using communication channels for navigation message transmission~\cite{INAC1}. Moreover, the integration of communication and navigation services via near-space platforms has shown distinct advantages, particularly in wide-range monitoring applications, outperforming traditional satellite and airborne networks~\cite{INAC2}. A crossover multiple-way ranging protocol for device-to-device positioning measure in 5G communication networks was proposed without increasing resource demands, compared to traditional two-way ranging protocols~\cite{INAC3}. As mentioned above, the conventional INAC network usually relies on time-division-multiple-access techniques, where navigation and communication signals are separated by time resource blocks, potentially compromising navigation performance~\cite{TDMA_navi}. Therefore, obtaining a novel solution that can simultaneously provide communication and navigation services becomes the primary goal. Non-orthogonal multiple access (NOMA) is increasingly recognized as an effective technique in wireless communications, facilitating access for numerous users~\cite{NOMA0,NOMA1,NOMA2}. The NOMA technique can serve users at different quality-of-service (QoS) requirements, which is natural to implement the navigation and communication services in the same time/frequency/code resource blocks by NOMA technique for the INAC networks~\cite{INAC001}. The RIS-NOMA-aided INAC network was initially explored in~\cite{INAC_Hou1}, which examined both communication-oriented and navigation-oriented INAC. Key performance indicators were evaluated to determine the effectiveness of INAC networks, e.g., outage probability, channel capacity, and navigation accuracy.

Based on the previous contributions, the STAR-RIS-NOMA-aided INAC in the urban canyons has not been studied extensively, resulting in the following three additional challenges: i) Since the LEO network expects more than ten thousand satellites, the cost efficiency cannot satisfy the requirements of 6G. Consequently, this raises an important question: Can MEO satellites be utilized for integrated communication and navigation services? ii) To overcome the requirement of LoS links in the conventional GNSS, can STAR-RIS provide additional links for urban canyon and indoor scenarios? iii) Since multiple satellites are located in orbit simultaneously, the satellite selection algorithm providing the optimized positioning accuracy and communication performance is still expected. Therefore, this article aims at integrating STAR-RIS and NOMA techniques into the INAC networks, where the high data rate requirement of communications and the high positioning accuracy requirement of navigation in urban canyons can be both satisfied.

Based on the above background, our main contributions are concluded as follows:
\begin{itemize}
  \item  In this work, we introduce a pioneering STAR-RIS-NOMA-assisted MEO satellite INAC network specifically designed for urban canyons. By utilizing the superposition coding (SC) technique, navigation and communication signals are efficiently fused in the same time/frequency/code resource blocks. Subsequently, users employ the successive interference cancellation (SIC) technique to distinguish between navigation and communication signals, which effectively conserves the resources of satellites.
  \item We propose both CO-INAC and NO-INAC based on the power allocation factors of SC. We then formulate the positioning equations of the proposed STAR-RIS assisted INAC network, and the position dilution of precision (PDoP) is computed. Meanwhile, the achievable ergodic rate and the positioning accuracy are calculated to demonstrate the feasibility of STAR-RIS-aided INAC networks.
  \item The STAR-RIS array is deployed at the proper location that can provide LoS links for the users located in the urban canyon and indoor scenarios, e.g., on the window of buildings. We then proposed two novel algorithms, namely navigation-prioritized-algorithm and communication-prioritized-algorithm, to select the satellite with the optimized navigation or communication performance. By doing so, the passive beamforming is optimized by utilizing a penalty-based optimization algorithm.
  \item The simulation results demonstrate that: 1) the proposed STAR-RIS-NOMA-assisted INAC systems can provide revolutionary solutions in the urban canyon and indoor scenarios; 2) The indoor users suffer higher positioning error than the outdoor users, where the positioning error of the indoor users mainly depends on the distance between the STAR-RIS and users; 3) The nearest INAC satellite is expected to enhance the communication performance, which may not proper for the navigation services.
\end{itemize}

\subsection{Organization}
In Section \uppercase\expandafter{\romannumeral2}, the system model, signal model, and urban canyons are proposed. Section \uppercase\expandafter{\romannumeral3} describes the positioning principles of this article. In Section \uppercase\expandafter{\romannumeral4}, we introduce our novel objective functions and passive beamforming at STAR-RIS. The simulation analysis of the proposed STAR-RIS-NOMA-assisted INAC networks in terms of positioning accuracy and achievable ergodic rate are verified in Section \uppercase\expandafter{\romannumeral4}. Finally, section \uppercase\expandafter{\romannumeral5} concludes this paper.

\section{System Model}

\begin{figure}[ht]
\centering
\includegraphics[width =3.5in]{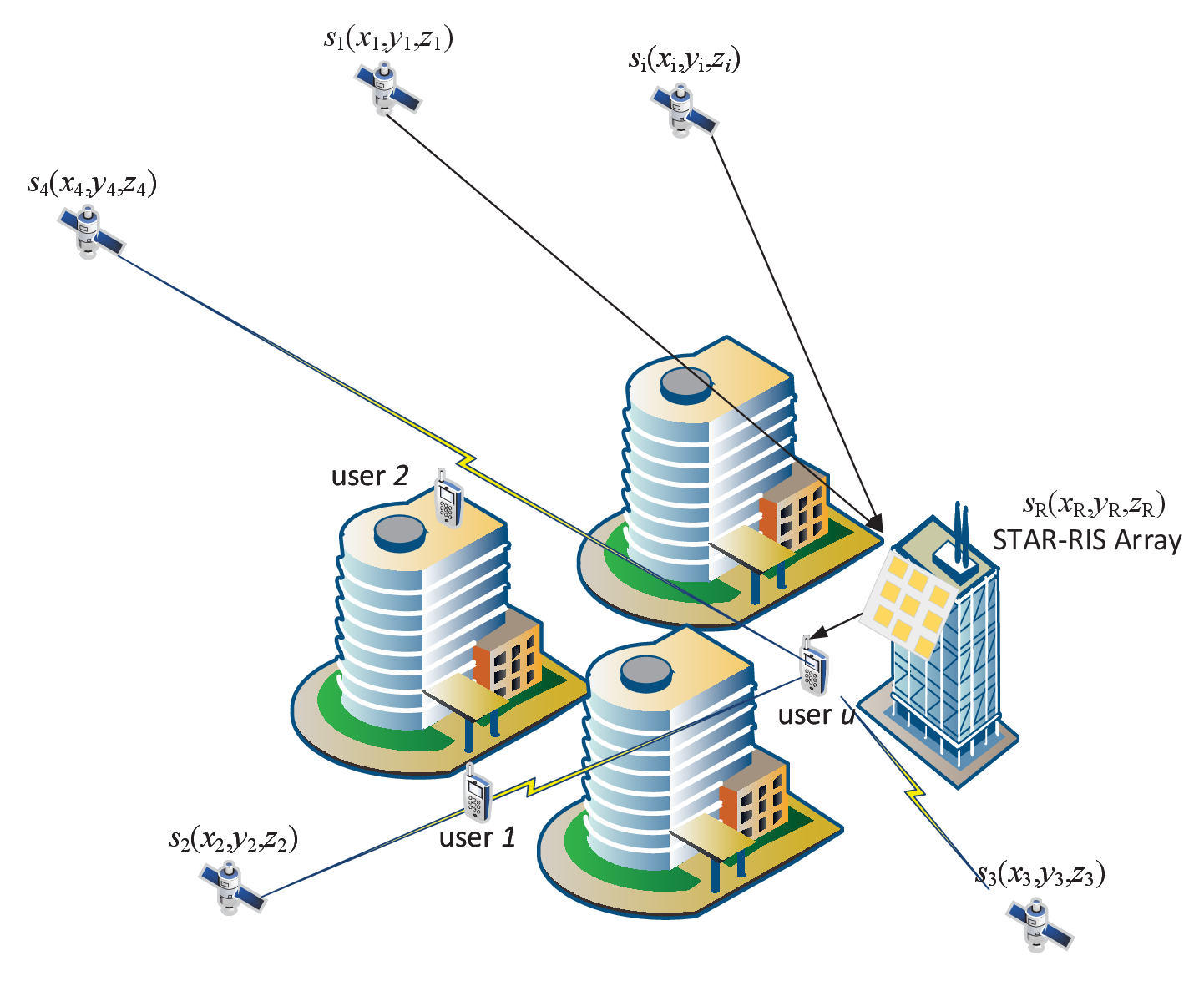}
\caption{The illustration of the STAR-RIS-NOMA-assisted INAC networks. }
\label{RISm}
\end{figure}

This article examines a new STAR-RIS-NOMA-aided INAC network as shown in Fig.~\ref{RISm}.
$M$ and $N$ denote the number of transmitting antennas (TAs) at the satellite and number of STAR-RIS elements, respectively.

%

\subsection{Channel Model}

The network contains $I$ satellites, $U$ users, and a STAR-RIS array. We first define the $i$-th satellite as $\varsigma_i$. Then, we pay our attention to the large-scale fading channels of $\varsigma_i - user_u$, which can be expressed as
\begin{equation}\label{large-scale_satell-user}
{L_{iu}} = {G_T}{\left( {\frac{\lambda }{{4\pi {d_{iu}}}}} \right)^2},
\end{equation}
where ${G_T}$ denote the transmit antenna gain. $\lambda $ denotes the signal wavelength. $d_{iu}$ is the distance of satellite $i$ and user $u$.

Note that in the urban canyons, the direct link between satellite and ground users may not exist. Therefore, the STAR-RISs can provide supplementary links between satellite and ground users, and the large scale fading channels of $\varsigma_i -STAR-RIS - user_u$ can be given by:
\begin{equation}\label{large-scale_satell-RIS}
{L_{Riu}} = {G_T}{G_R}{\left( {\frac{\lambda }{{4\pi {d_{iR}}}}} \right)^2}{\left( {\frac{\lambda }{{4\pi {d_{Ru}}}}} \right)^2},
\end{equation}
where $d_{iR}$ and $d_{Ru}$ respectively represent the distance of $\varsigma_i -STAR-RIS$ and $STAR-RIS - user_u$ links.

We then define the small-scale fading channel gain of $\varsigma_i - user_u$ link, ${h_{iu}}$, which follows the shadowed Rician fading. Although $M$ TAs are deployed, based on the fixed beamforming strategy adopted in GNSS and Beidou system, the multi-antenna gain can be considered as a single variable. Therefore, the small-scale fading vector ${{\bf{h}}_{u}}$ received by user $u$ can be expressed as
\begin{equation}\label{small-scale-channel-satell-user}
{{\bf{h}}_u} = \left[ {\begin{array}{*{20}{c}}
{{h_{1u}}}\\
{{h_{2u}}}\\
 \vdots \\
{{h_{iu}}}
\end{array}} \right],
\end{equation}
where ${{\bf{h}}_{u}}$ is a $\left( {i \times 1} \right)$-element vector.
Note that when the direct link between satellite $i$ and user $u$ does not exist, the corresponded row in~\eqref{small-scale-channel-satell-user} can be simply set to 0. Each element follows shadowed Rician fading channel gains with parameters $\Omega$, $2b$ and $m$, where $\Omega$ and $2b$ denote the average power of the LoS component and the multipath component, respectively. $m$ is the Nakagami-$m$ parameter ranging
from zero to infinite.

The small-scale channel matrix of $\varsigma_i -STAR-RIS$ links is expressed as
\begin{equation}\label{small_scale_satel_RIS_column_i}
{{\bf{h}}_{Ri}} = \left[ {\begin{array}{*{20}{c}}
{\begin{array}{*{20}{c}}
{\begin{array}{*{20}{c}}
{{h_{R,i1}}}\\
{{h_{R,i2}}}
\end{array}}\\
 \vdots
\end{array}}\\
{{h_{R,iN}}}
\end{array}} \right],
\end{equation}
where ${{h_{R,in}}}$ denotes the small-scale fading gain between satellite $i$ and the $n$-th STAR-RIS element, which follows shadowed Rician fading gains.

Similarly, the small-scale channel matrix of $STAR-RIS - user_u$ is given by:
\begin{equation}\label{small_scale_matrix_RIS_user_row_j}
{{\bf{g}}_u} = \left[ {\begin{array}{*{20}{c}}
{{g_{R,1u}}}& \cdots &{{g_{R,Nu}}}
\end{array}} \right],
\end{equation}
where ${{g_{R,nu}}}$ denotes the small-scale channel gain between the $n$-th STAR-RIS element and user $u$, which follows the Rician fading channel gains.

The STAR-RIS adopts both the reflective and transmit layers, thus we define the reflection and transmission matrix as:
\begin{equation}\label{reflection coefficient}
{{\bf{\Psi }}_R}{\rm{ = diag}}\left[ {\begin{array}{*{20}{c}}
{{\beta _{1,R}}{\phi _{1,R}}}&{{\beta _{2,R}}{\phi _{2,R}}}& \cdots &{{\beta _{N,R}}{\phi _{N,R}}}
\end{array}} \right],
\end{equation}
and
\begin{equation}\label{transmission coefficient}
{{\bf{\Psi }}_T}{\rm{ = diag}}\left[ {\begin{array}{*{20}{c}}
{{\beta _{1,T}}{\phi _{1,T}}}&{{\beta _{2,T}}{\phi _{2,T}}}& \cdots &{{\beta _{N,T}}{\phi _{N,T}}}
\end{array}} \right],
\end{equation}
where $j = \sqrt { - 1} $, ${\phi _{n,R}} = \exp (j{\theta _{n,R}})$ and ${\phi _{n,T}} = \exp (j{\theta _{n,T}})$ respectively represent the phase shift of the reflected and transmitted signals of STAR-RIS element ${n}$. $ 0 \le {\theta _{n,R}} < 2\pi  $, $0 \le {\theta _{n,T}} < 2\pi $. $0 < {\beta _{n,R}} \le 1$ and $0 < {\beta _{n,T}} \le 1$ denote the reflection amplitude coefficient and transmission amplitude coefficient of the ${n}$-th STAR-RIS element with $\beta _{n,R}^2 + \beta _{n,T}^2 = 1$, respectively.

Regarding ground user $u$ located in the urban canyons, without loss of generality, it is assumed that the number of visible satellites for user $u$ is defined as ${I_v}, {I_v} \in I$.
In addition, some satellites are invisible for user $u$, which are defined by a set of ${I_n},{I_n} \in I$.
Hence, if user $u$ is located at the reflection space of STAR-RIS, the received signals of user $u$ can be expressed as
\begin{equation}\label{received_signal_reflected_user}
\begin{aligned}
{y_{u,R}}(t) = & \underbrace {\sum\limits_{i = 1,i \in {I_n}}^I {\left( {{{\bf{g}}_u}{{\bf{\Psi }}_R}{{\bf{h}}_{Ri}}} \right)\sqrt {{L_{Riu}}{p_i}} {s_i}(t)} }_{{\rm{Reflected~Signals}}} \\
 &+ \underbrace {\sum\limits_{i = 1,i \in {I_v}}^I {{h_{iu}}} \sqrt {{L_{iu}}{p_i}} {s_i}(t)}_{{\rm{Directed~Signals}}} + {n_0},
\end{aligned}
\end{equation}
where ${n_0}$ denotes the additive white Gaussian noise (AWGN), ${s_i}(t)$ denotes the information of the $i$-th satellite. ${p_i}$ represents the transmit power of satellites $i$.

On the other hand, if user $u$ is located in the indoor scenarios, the transmitted signals can be expressed as
\begin{equation}\label{received_signal_transsmited_user}
{y_{u,T}}(t) = \underbrace {\sum\limits_{i = 1,i \in {I_n}}^I {\left( {{{\bf{g}}_u}{{\bf{\Psi }}_T}{{\bf{h}}_{Ri}}} \right)\sqrt {{L_{Riu}}{p_i}} {s_i}(t)} }_{{\rm{Transmitted~Signals}}} + {n_0}.
\end{equation}

\subsection{Signal Model}

Due to the fact that the code division technique is deployed in satellite navigation, the users can separate the information of different satellites. Therefore, we simply assume that one INAC satellite is deployed. The SC and SIC techniques are employed at the INAC satellite and ground users to detect the composite signals, respectively. Then, the INAC satellite sends the following signal:
\begin{equation}\label{NOMA model}
{s}(t) = {\omega _N}{s_N}(t) + {\omega _C}{s_C}(t),
\end{equation}
where ${\omega _N}$ and ${\omega _C}$ represent the power allocation factors of navigation and communication signals, respectively. ${s_N}(t)$ and ${s_C}(t)$ denote navigation and communication signals, respectively. Under the NOMA protocol, the power allocation factors are constrained such that $\omega_N^2 + \omega_C^2 = 1$. As can be seen from~\eqref{NOMA model}, two potential cases with different power allocation factors, NO-INAC and CO-INAC, can be investigated.

\subsubsection{NO-INAC Case}
It is assumed that the power allocation factors satisfy
\begin{equation}
{\omega _C}>{\omega _N} .
\end{equation}

In this scenario, the INAC user initially decodes the communication signal, treating the navigation signal as interference, and the signal-to-interference-plus-noise ratio (SINR) can be expressed as follows:
\begin{equation}\label{snrn_NO_communication}
SINR_C^{NO} = \frac{{{{\left| {{{\tilde h}_u}} \right|}^2}\omega _C^2{p_i}}}{{{{\left| {{{\tilde h}_u}} \right|}^2}\omega _N^2{p_i} + {\sigma ^2}}},
\end{equation}
where ${\left| {{{\tilde h}_u}} \right|^2}$ denotes the channel gain of user $u$, ${{\sigma ^{\rm{2}}}}$ represents the power of AWGN.
On the one hand, when both the direct link and reflect link exist, the channel gain can be expressed as
\begin{equation}\label{channel gain both direct and reflect link}
{\left| {{{\tilde h}_u}} \right|^2} = {\left| {{{\bf{g}}_u}{{\bf{\Psi }}_R}{{\bf{h}}_{Ri}}\sqrt{{L_{Riu}}}} +{{h_{iu}}\sqrt{{L_{iu}}}} \right|^2}.
\end{equation}
On the other hand, when only the reflect link exists, the effective channel gain is given by
\begin{equation}\label{channel gain only refect link}
{\left| {{{\tilde h}_u}} \right|^2} = {\left| {{{\bf{g}}_u}{{\bf{\Psi }}_R}{{\bf{h}}_{Ri}}\sqrt{{L_{Riu}}}} \right|^2}.
\end{equation}

After the communication signals are subtracted from the received signals, the navigation signals can be detected, then the SINR of navigation signals is given by
\begin{equation}\label{snrc0}
SINR_N^{NO} = \frac{{{{\left| {{{\tilde h}_u}} \right|}^2}\omega _N^2{p_i}}}{{{\sigma ^{\rm{2}}}}}.
\end{equation}

Therefore, in the NO-INAC case, the achievable navigation rate and communication rate of the INAC user ca be given by
\begin{equation}\label{NO_navigation_rate}
{R^{NO}_N} = {\log _2}\left( {1 + SINR_N^{NO}} \right),
\end{equation}
and
\begin{equation}\label{NO_communication_rate}
{R^{NO}_C} = {\log _2}\left( {1 + SINR_C^{NO}} \right),
\end{equation}
respectively.

\subsubsection{CO-INAC Case}
In the CO-INAC scenarios, we prioritize communication services, leading to the condition where ${\omega _N}$ and ${\omega _C}$ satisfy ${\omega _C} < {\omega _N}$. Thus, following the NOMA protocol, navigation signals should be decoded prior to communication signals. Consequently, the SINR for decoding navigation signals can be formulated as:
\begin{equation}\label{snrn_CO_N}
SINR_N^{CO} = \frac{{{{\left| {{{\tilde h}_u}} \right|}^2}\omega _N^2{p_i}}}{{{{\left| {{{\tilde h}_u}} \right|}^2}\omega _C^2{p_i} + {\sigma ^2}}}.
\end{equation}

After the navigation signals are deleted from the received signal, the communication signals can be obtained with the following SINR
\begin{equation}\label{snrc1}
SINR_C^{CO} = \frac{{{{\left| {{{\tilde h}_u}} \right|}^2}\omega _C^2{p_i}}}{{{\sigma ^{\rm{2}}}}}.
\end{equation}

Therefore, the achievable rates of communication and navigation of users in the CO-INAC cases are given by
\begin{equation}\label{rate CO_C}
{R^{CO}_C} = {\log _2}\left( {1 + SINR_C^{CO}} \right),
\end{equation}
and
\begin{equation}
{R^{CO}_N} = {\log _2}\left( {1 + SINR_N^{CO}} \right),
\end{equation}
respectively.

\section{Problem Formulation}

\subsection{Geometric Positioning Principle}
In this subsection, we first describe the geometric positioning principle.
The locations of user $u$, satellite ${s_i}$ and STAR-RIS array are specified as $({x_u},{y_u},{z_u})$, $({x_i},{y_i},{z_i})$ and $({x_R},{y_R},{z_R})$, respectively.

The actual distance of $\varsigma_i - user_u$ link is denoted by ${r_{\tau ,iu}}$, where $i = 1, 2, \cdots, I$. The distances of $\varsigma_i -STAR-RIS$ and $STAR-RIS - user_u$ links are represented as ${r_{\tau iR}}$ and ${r_{\tau Ru}}$, respectively. The GNSS system utilizes a highly accurate clock system that precisely records the transmission time ${t_{si}}$ when satellite ${s_i}$ sends a message frame. Similarly, user $u$ records the reception time ${t_{ru}}$ of the message. The distance ${r_{\tau ,iu}}$ between satellite ${s_i}$ and user $u$ is then calculated using the speed of light. Likewise, by employing both the speed of light and the transmission times, the combined distance through the STAR-RIS array, expressed as ${r_{\tau iR}} + {r_{\tau Ru}}$, is also computed.

When a direct link between satellite ${s_i}$ and the ground user is established, the positioning equation for satellite ${s_i}$ can be derived from the Euclidean distance as follows:
\begin{equation}\label{e13}
{r_{\tau ,iu}} = \sqrt {{{({x_i} - {x_u})}^2} + {{({y_i} - {y_u})}^2}{\rm{ + }}{{({z_i} - {z_u})}^2}} ,
\end{equation}
where ${r_{\tau ,iu}}$ represents the distance of $\varsigma_i - user_u$ link, which can be estimated by pseudo-range.

\subsection{Pseudo-Range Positioning}


In practical applications, the pseudo-range primarily comprises clock error and atmospheric delay error. It is assumed that satellite ${s_i}$ transmits a navigation message at its clock time ${t_{si}}$, and user $u$ receives this message at their local time ${t_{ru}}$. Thus, the estimated distance of $\varsigma_i - user_u$ link can be expressed as:
\begin{equation}\label{tranmit time}
{\rho _{t,iu}} = c({t_{ru}} - {t_{si}}).
\end{equation}

Given that the satellite clock is not perfectly synchronized with the standard time, it is assumed that the clock error between satellite ${s_i}$ and the standard clock to be $\Delta{t_{si}}$, and the clock error between user $u$'s clock and the standard clock to be $\Delta{t_{ru}}$. Additionally, the delays caused by the ionosphere and troposphere are represented as ${I_i}$ and ${T_i}$, respectively. Other systematic errors are $\Delta{\varepsilon _i}$. Considering all the above factors, the actual distance of $\varsigma_i - user_u$ can be further transformed into:
\begin{equation}\label{distance satellite user}
{r_{\tau ,iu}} = {\rho _{t,iu}} - c(\Delta {t_{ru}} - \Delta {t_{si}} + {I_i} + {T_i} + \Delta {\varepsilon _i}),
\end{equation}
where $\Delta {t_{si}}$ can be obtained by the navigation messages. Thus, the pseudo-range can be simplified to
\begin{equation}
{\rho _{ci}} = {\rho _{t,iu}} + c(\Delta {t_{si}} - {I_i} - {T_i}).
\end{equation}

By doing so, the actual distance of $\varsigma_i - user_u$ can be simplified to
\begin{equation}\label{e119}
{r_{\tau ,iu}} = {\rho _{ci}} - c\Delta {t_r}.
\end{equation}

Based on~\eqref{e13} to~\eqref{e119}, the goal of this article is to calculate four unknowns ${x_u}$, ${y_u}$, ${z_u}$, and ${\Delta{t_r}}$. Thus, we assume that user $u$ receives several signals from satellite, which are given by
\begin{equation}\label{e20}
\begin{aligned}
 \sqrt {{{\left( {{x_1} - {x_u}} \right)}^2} + {{\left( {{y_1} - {y_u}} \right)}^2} + {{\left( {{z_1} - {z_u}} \right)}^2}}  &= {\rho _{c1}} - c\Delta {t_r}, \\
&  \vdots \\
\sqrt {{{\left( {{x_i} - {x_u}} \right)}^2} + {{\left( {{y_i} - {y_u}} \right)}^2} + {{\left( {{z_i} - {z_u}} \right)}^2}}  &  = {\rho _{ci}} - c\Delta {t_r}.
\end{aligned}
\end{equation}

Note that in practice, at least four satellites are required for navigation positioning. However, in the urban canyons, the direct link between satellite and user may be blocked. Fortunately, the STAR-RIS is able to provide extended LoS path between satellite and user. Therefore, the positioning equation between satellite $v$ and user $u$ through STAR-RIS is given by
\begin{equation}\label{e191}
{r_{\tau ,vR}} + \sqrt {{{({x_R} - {x_u})}^2} + {{({y_R} - {y_u})}^2}{\rm{ + }}{{({z_R} - {z_u})}^2}} {\rm{ = }}{\rho _{cv}} - c\Delta {t_r},
\end{equation}
where ${\rho _{cv}}$ represents the modified pseudo-range between the $v$-th satellite to user $u$ via STAR-RIS. The Euclidean distance of the $v$-th satellite to STAR-RIS is defined by ${r_{\tau ,vR}}$ as:
\begin{equation}\label{e21}
{r_{\tau ,vR}} = \sqrt {{{({x_4} - {x_R})}^2} + {{({y_4} - {y_R})}^2}{\rm{ + }}{{({z_4} - {z_R})}^2}} .
\end{equation}

We can re-formulate multiple positioning equations as follows:
\begin{equation}\label{e30}
\begin{aligned}
{\rho _{c1}}\left( {{{\bf{x}}_r}} \right) & = \sqrt {{{\left( {{x_1} - {x_u}} \right)}^2} + {{\left( {{y_1} - {y_u}} \right)}^2} + {{\left( {{z_1} - {z_u}} \right)}^2}}  + c\Delta {t_r},\\
 & \vdots \\
{\rho _{ci}}\left( {{{\bf{x}}_r}} \right) & = \sqrt {{{\left( {{x_i} - {x_u}} \right)}^2} + {{\left( {{y_i} - {y_u}} \right)}^2} + {{\left( {{z_i} - {z_u}} \right)}^2}}  + c\Delta {t_r},\\
{\rho _{cv}}\left( {{{\bf{x}}_r}} \right) & = {r_{\tau ,vR}} + \sqrt {{{({x_R} - {x_u})}^2} + {{({y_R} - {y_u})}^2}{\rm{ + }}{{({z_R} - {z_u})}^2}} \; \\
&+ c\Delta {t_r},
\end{aligned}
\end{equation}
where ${{\bf{x}}_r} = {[{x_u},{y_u},{z_u},\Delta{t_r}]^T}$, ${\rho _{ci}}({{\bf{x}}_r})$ represent the corrected pseudo-range of $\varsigma_i - user_u$ link.

We adopt the least squares method (LSM) algorithm to evaluate the position of users, and the state equation is given by
\begin{equation}\label{equa0}
{\bf{b}} = {\bf{A}}{d}{\bf{x}} + {N_0},
\end{equation}
where \textbf{x} denotes a vector with $k \times 1$ unknowns, \textbf{b} represents an ${\rm{4}} \times {\rm{1}}$ vector.
${\bf{A}}$ is a $\left( {i + v} \right) \times {\rm{4}}$ matrix, which can be calculated from the navigation messages as follows:
\begin{equation}\label{V}
{\bf{A}} = {\left[ {\begin{array}{*{20}{c}}
{{{\bf{u}}_1}}& \cdots &{{{\bf{u}}_i}}&{{{\bf{u}}_v}}
\end{array}} \right]^T},
\end{equation}
where ${{\bf{u}}_i} = \left[ {\begin{array}{*{20}{c}}
{\frac{{{x_i} - {x_{u0}}}}{{{r_i}\left( {{{\bf{x}}_{r0}}} \right)}}}&{\frac{{{y_i} - {y_{u0}}}}{{{r_i}\left( {{{\bf{x}}_{r0}}} \right)}}}&{\frac{{{z_i} - {z_{u0}}}}{{{r_i}\left( {{{\bf{x}}_{r0}}} \right)}}}&c
\end{array}} \right]$, and $ d{{\bf{x}}_0} = \left[ {\begin{array}{*{20}{c}}
{{x_u} - {x_{u0}}}\\
{{y_u} - {y_{u0}}}\\
{{z_u} - {z_{u0}}}\\
{\left( {{\rm{\Delta }}{t_r} - {\rm{\Delta }}{t_{r0}}} \right)}
\end{array}} \right]$.

The cost function ${\bf{P}}({\bf{x}})$ is defined as
\begin{equation}\label{P(X)}
\begin{aligned}
\mathbf{P}(\mathbf{x}) &=(\mathbf{A x}-\mathbf{b})^{T}(\mathbf{A} \mathbf{x}-\mathbf{b}) \\
& = {{\bf{x}}^{\rm{T}}}{{\bf{A}}^{\rm{T}}}{\bf{Ax}} - 2{{\bf{x}}^{\rm{T}}}{{\bf{A}}^{\rm{T}}}{\bf{b}} + {{\bf{b}}^{\rm{T}}}{\bf{b}},
\end{aligned}
\end{equation}
where ${\mathbf{P}(\mathbf{x})}$ is the sum square error. The goal of LSM is to minimize the sum of squared errors by differentiating ${\mathbf{P}(\mathbf{x})}$ with respect to $x$, we can derive:

\begin{equation}\label{derivative of P(x)}
\frac{{\partial {\bf{P}}({\bf{x}})}}{{\partial {\bf{x}}}} = 2{{\bf{A}}^{\rm{T}}}{\bf{Ax}} - 2{{\bf{A}}^{\rm{T}}}{\bf{b}}.
\end{equation}

The extreme point of the cost function is
\begin{equation}\label{extreme point of P(x)}
{\bf{\hat x}} = {\left( {{{\bf{A}}^{\rm{T}}}{\bf{A}}} \right)^{ - 1}}{{\bf{A}}^{\rm{T}}}{\bf{b}}.
\end{equation}

Since the second derivative of ${\mathbf{P}(\mathbf{x})}$ with respect to $x$ $\frac{{{\partial ^2}{\bf{P}}({\bf{\hat x}})}}{{\partial {{{\bf{\hat x}}}^2}}} = 2{{\bf{A}}^{\rm{T}}}{\bf{A}}$ is positive definite, ${\mathbf{P}(\mathbf{x})}$ is minimized when the derivative is 0. Then the estimation error $\delta {\bf{x}}$ is defined as
\begin{equation}\label{delta x}
\begin{aligned}
\delta {\bf{x}}  &=\mathbf{x}-\left(\mathbf{A}^{\rm{T}} \mathbf{A}\right)^{-1} \mathbf{A}^{\rm{T}} \mathbf{b} \\
& ={\bf{x}}-\left(\mathbf{A}^{\rm{T}} \mathbf{A}\right)^{-1} \mathbf{A}^{\rm{T}}(\mathbf{A} {\bf{x}}+n) \\
& =-\left(\mathbf{A}^{\rm{T}} \mathbf{A}\right)^{-1} \mathbf{A}^{\rm{T}} \mathbf{n}.
\end{aligned}
\end{equation}

According to (\ref{delta x}), the least squares estimation is only related to the system transformation relation vector A and the system noise $N$. When  $N$ follows a Gaussian distribution with a mean of 0, the estimation of LSM is unbiased. Since LSM can only solve linear equations, it is necessary to transform the problem into a linear one.

We then define that ${{\rho _{ci}}\left( {{{\bf{x}}_{r0}}} \right)}$ denotes the corrected pseudo-range between satellite ${s_i}$ and initial solution, which is expressed as
\begin{equation}
\begin{array}{l}
{\rho _{ci}}\left( {{{\bf{x}}_{r0}}} \right) = \sqrt {{{\left( {{x_i} - {x_{u0}}} \right)}^2} + {{\left( {{y_i} - {y_{u0}}} \right)}^2} + {{\left( {{z_i} - {z_{u0}}} \right)}^2}} + c{\rm{\Delta }}{t_{r0}}\\
\ \ \ \ \ \ \  \ \ \ \ {\rm{ = }}{r_i}\left( {{{\bf{x}}_{r0}}} \right) + c{\rm{\Delta }}{t_{r0}},
\end{array}
\end{equation}
where ${r_i}\left( {{{\bf{x}}_{r0}}} \right)$ represents the Euclidean distance from satellite ${s_i}$ to initial solution.

Then, the positioning equations is transformed into
\begin{equation}\label{e31}
\begin{aligned}
& \left[ {\begin{array}{*{20}{c}}
{{\rho _{c1}}({{\bf{x}}_r}) - {\rho _{c1}}({{\bf{x}}_{r0}})}\\
{{\rho _{c2}}({{\bf{x}}_r}) - {\rho _{c2}}({{\bf{x}}_{r0}})}\\
{{\rho _{c3}}({{\bf{x}}_r}) - {\rho _{c3}}({{\bf{x}}_{r0}})}\\
{{\rho _{c4}}({{\bf{x}}_r}) - {\rho _{c4}}({{\bf{x}}_{r0}})}
\end{array}} \right]\; =\\
& \left[ {\begin{array}{*{20}{c}}
{\frac{{{x_1} - {x_{r0}}}}{{{r_1}\left( {{{\bf{x}}_{r0}}} \right)}}}&{\frac{{{y_1} - {y_{r0}}}}{{{r_1}\left( {{{\bf{x}}_{r0}}} \right)}}}&{\frac{{{z_1} - {z_{r0}}}}{{{r_1}\left( {{{\bf{x}}_{r0}}} \right)}}}&c\\
{\frac{{{x_2} - {x_{r0}}}}{{{r_2}\left( {{{\bf{x}}_{r0}}} \right)}}}&{\frac{{{y_2} - {y_{r0}}}}{{{r_2}\left( {{{\bf{x}}_{r0}}} \right)}}}&{\frac{{{z_2} - {z_{r0}}}}{{{r_2}\left( {{{\bf{x}}_{r0}}} \right)}}}&c\\
{\frac{{{x_3} - {x_{r0}}}}{{{r_3}\left( {{{\bf{x}}_{r0}}} \right)}}}&{\frac{{{y_3} - {y_{r0}}}}{{{r_3}\left( {{{\bf{x}}_{r0}}} \right)}}}&{\frac{{{z_3} - {z_{r0}}}}{{{r_3}\left( {{{\bf{x}}_{r0}}} \right)}}}&c\\
{\frac{{{x_4} - {x_{r0}}}}{{{r_4}\left( {{{\bf{x}}_{r0}}} \right)}}}&{\frac{{{y_4} - {y_{r0}}}}{{{r_4}\left( {{{\bf{x}}_{r0}}} \right)}}}&{\frac{{{z_4} - {z_{r0}}}}{{{r_4}\left( {{{\bf{x}}_{r0}}} \right)}}}&c
\end{array}} \right]\left[ {\begin{array}{*{20}{c}}
{{x_u} - {x_{r0}}}\\
{{y_u} - {y_{r0}}}\\
{{z_u} - {z_{r0}}}\\
{\Delta {t_r} - \Delta {t_{r0}}}
\end{array}} \right] \\
&+ \left[ {\begin{array}{*{20}{c}}
{{n_1}}\\
{{n_2}}\\
{{n_3}}\\
{{n_4}}
\end{array}} \right].\;
\end{aligned}
\end{equation}

Hence, we can calculate the positioning for satellite ${s_i}$ by the following equation
\begin{equation}\label{equa2}
\delta {\rho _{ci}} = {{\bf{u}}_i}{({{\bf{A}}^T}{\bf{A}})^{ - 1}}{{\bf{A}}^T}{\bf{b}} + {n_0}.
\end{equation}

The current solution ${{\bf{x}}_{r1}}$ can be then updated, which can be expressed as
\begin{equation}
{{\bf{x}}_{r1}} = {{\bf{x}}_{r0}} + {({{\bf{A}}^T}{\bf{A}})^{ - 1}}{{\bf{A}}^T}{\bf{b}}.
\end{equation}

%
%
%

\subsection{Positioning accuracy}

We first define a positioning error and clock error vector, which can be calculated from the pseudo-range measurement error as
\begin{equation}
{\left[ {\begin{array}{*{20}{c}}
{{\varepsilon _x}}&{{\varepsilon _y}}&{{\varepsilon _z}}&{{\varepsilon _t}}
\end{array}} \right]^T} = {\left( {{{\bf{A}}^T}{\bf{A}}} \right)^{ - 1}}{{\bf{A}}^T}{\varepsilon _\rho },
\end{equation}
where ${{\bf{\varepsilon }}_{\bf{\rho }}}$ denotes the measurement error, which is independent and follows Gaussian distributions with mean 0 and variance $\sigma _{URE}^2$. In order to provide comprehensive performance analysis, we first derived the covariance of pseudo-range measurement error as
\begin{equation}\label{equa4}
\begin{array}{l}
Cov\left( {{{\left[ {\begin{array}{*{20}{c}}
{{\varepsilon _x}}&{{\varepsilon _y}}&{{\varepsilon _z}}&{{\varepsilon _t}}
\end{array}} \right]}^T}} \right)\;\;\\
 = {\left( {{{\bf{A}}^T}{\bf{A}}} \right)^{ - 1}}\sigma _{URE}^2\\
 = {\bf{F}}\sigma _{URE}^2,
\end{array}
\end{equation}
where $Cov\left( {\rm{\cdot}} \right)$ denotes the covariance function. ${\bf{F}} = {({{\bf{A}}^T}{\bf{A}})^{ - 1}}$ is the weight coefficient matrix, which is a fourth-order symmetric matrix.

It is clear that the measurement error is amplified by the weight coefficient matrix ${\bf{F}}$ and becomes the positioning error.
$\sigma _x^2$, $\sigma _y^2$, $\sigma _z^2$ and $\sigma _t^2$ are the variances of positioning error, which correspond to the diagonal elements ${h_{11}}$, ${h_{22}}$, ${h_{33}}$ and ${h_{44}}$, respectively. In this article, the PDoP is utilized to reveal the performance of the proposed networks, which is formulated as follows:
\begin{equation}\label{GDoP}
\begin{aligned}
{\rm{PDoP}} &= \sqrt {\sigma _x^2 + \sigma _y^2 + \sigma _z^2 }\\ &= \sqrt {{h_{11}} + {h_{22}} + {h_{33}} }.
\end{aligned}
\end{equation}
Note that based on the definition of PDoP, when the satellites are evenly distributed, PDoP becomes smaller, resulting in higher position accuracy.

\section{Passive Beamforming Design at RIS}

In this article, we consider that the number of visible satellites is lower than 4. The STAR-RIS array is deployed to provide navigation and communication services to the users located in the urban canyon or indoor scenarios. Therefore, the STAR-RIS needs to select a proper satellite based on the different priorities mentioned in the following subsections.

\subsubsection{Navigation-Prioritized Algorithm (NPA)}

In the urban canyons, the STAR-RIS array reflects the invisible satellites to the ground users to provide additional links. In order to provide higher accurate navigation services, we propose a novel NPA based on the PDoP by designing the passive beamforming at STAR-RIS, where the navigation accuracy of multiple users is optimized. To do so, we first define the position error as follows:
\begin{equation}\label{position error of user u}
{\varepsilon _u} = \sqrt {{{\left( {x - {x_u}} \right)}^2} + {{\left( {y - {y_u}} \right)}^2} + {{\left( {z - {z_u}} \right)}^2}},
\end{equation}
where $x$, $y$ and $z$ are calculated position of user $u$. ${x_u}$, ${y_u}$ and ${z_u}$ are real position of user $u$.

In order to improve the positioning accuracy, the objective function can be given by:
\begin{equation}\label{objective function NPA}
\begin{array}{l}
\begin{aligned}
\min_{{{s}_{sel}}} & \; \varepsilon _u \left( x,y,z \right) \\
{\rm{subject}}\;{\rm{to}}\; & {\bf{\hat x}}<\epsilon \\
&\mathrm{n}<\mathrm{n}^{*}\\
&{{s}_{sel}}\in {{I}_{n}},
\end{aligned}
\end{array}
\end{equation}
where $\mathrm{n}^{*}$ is the maximum number of iterations of the least squares algorithm. ${\bf{\hat x}}$ is the correction of the least squares estimate. ${\epsilon}$ is the pre-set iteration threshold. ${s_{sel}}$ is the satellite selected through RIS in the satellite selection method.

We then define that ${F({ \bar{\bullet}})}$ is the distance from the satellite to the user. ${\bf \bar{l}}$ is the pseudo-range matrix. ${{\bf C}_{\ell}}$ is the covariance matrix of measurements. ${{\bf{x}}^0}$ and  ${\bf{\bar x}}$ respectively are the initial value of the iteration and the user position guessed based on the pseudo-range. ${{\bf{x}}^{ls}}$ is the least squares estimate. Then, in order to select the satellite providing the best positioning accuracy, a penalty-based algorithm can be given by {\textbf{Algorithm~\ref{Algorithm 1}} as follows.
\begin{algorithm}[t]
\caption{Penalty-based iterative algorithm to solve problem~\eqref{objective function NPA}} \label{Algorithm 1}
\begin{algorithmic}[1]
    \REPEAT
    \STATE{In the selecting loop: obtain the PDoP of each NLoS satellite combined with the LoS satellites }
        \REPEAT
        \STATE {In the LSM loop:  Perform the least squares method}
        \STATE {Establish ${F({\bf \bar{x}})={\bf \bar{l}}}$}\\
        \STATE {Initialize ${{{\bf{x}}^0} \cong {\bf{\bar x}}}$ and ${\bf{Ax}} = {\bf{b}} + {\bf{v}}$\\
        \STATE initialize the iteration index $\rm n = 0$, solve $ {{{\bf{A}}^{\rm{T}}{\bf{A}}}}{\bf{\hat x}} = {{\bf{A}}^{\rm{T}}}{\bf{b}}$}
        \STATE{${\bf{\hat x}} = {\left( {{{\bf{A}}^{\rm{T}}}{\bf{A}}} \right)^{ - 1}}{{\bf{A}}^{\rm{T}}}{\bf{b}}$}
        \STATE{${{\bf{x}}^{ls}} = {{\bf{x}}^0} + {\bf{\hat x}}$}
        \UNTIL {${\bf{\hat x}}$ is below ${\epsilon}$}.
    \UNTIL{all the NLoS satellites have been traversed.}
    \STATE Select the NLoS satellite which minimizes the PDoP.
\end{algorithmic}
\end{algorithm}

With the aid of {\textbf{Algorithm~\ref{Algorithm 1}}, the satellite providing the best positioning performance can be selected by the NPA. Then, we can have following remarks.

\begin{remark}\label{remark1: Indoor share path}
The results in~\textbf{Algorithm~\ref{Algorithm 1}} demonstrate that although STAR-RIS can provide additional LoS links from satellite to users, a STAR-RIS array can only provide users with one virtual LoS link, that is, each STAR-RIS array can only introduce a maximum of one non-directly connected satellite. Therefore, the STAR-RIS to indoor user link is shared by multiple satellites, resulting in high positioning errors.
\end{remark}

\begin{remark}\label{remark2: navigation satellite expectation}
The results in~\textbf{Algorithm~\ref{Algorithm 1}} demonstrate that in order to improve the PDoP performance, the satellite with higher geometric diversity is expected for the NPA.
\end{remark}

Based on~\textbf{Algorithm~\ref{Algorithm 1}}, the NPA exhibits dual applications. Primarily, it facilitates simultaneous communication and navigation functionalities for users encountering a scarcity of directly accessible satellites. Secondly, it serves to supplement satellite provision, thereby enhancing positional accuracy for users experiencing low precision in positioning.

\subsubsection{Communication-Prioritized Algorithm (CPA)}

On the contrary, we also focuses on the communication performance of both the indoor and outdoor users. It is assumed that the INAC satellite is the $i$-th satellite. Therefore, we then propose a CPA for maximizing the achievable ergodic rate of users, where the objective function of CPA can be given by:

\begin{equation}\label{objective function CPA}
\begin{array}{l}
\begin{aligned}
\max_{{\theta _{1,R}}, {\theta _{1,T}}  } & \; R_R+R_T \\
{\rm{subject}}\;{\rm{to}}\; & {\beta _{1,R}} \cdots {\beta _{N,R}} = 0.5\\
& {\beta _{1,T}} \cdots {\beta _{N,T}} = 0.5\\
& {\theta _{1,R}} \cdots {\theta _{N,R}} \in \left[ {0,2\pi } \right)\\
& {\theta _{1,T}} \cdots {\theta _{N,T}} \in \left[ {0,2\pi } \right),
\end{aligned}
\end{array}
\end{equation}
where $R_R$ and $R_T$ respectively represent the achievable ergodic rate of the reflected and indoor users.

Given that navigation data requirements are substantially lower than that of communications, the passive beamforming at the STAR-RIS is designed primarily to optimize channel gains for both indoor and outdoor users. Consequently, the objective functions can be re-formulated as follows:
 \begin{equation}\label{objective function CPA_reformulate}
\begin{array}{l}
\begin{aligned}
\max_{{\theta _{1,R}}  } & {\left| {{{\bf{g}}_u}{{\bf{\Psi }}_R}{{\bf{h}}_R}\sqrt {{L_{Riu}}}  + {{\bf{h}}_{iu}}\sqrt {{L_{iu}}} } \right|^2}\\
\max_{ {\theta _{1,T}}  } & {\left| {{{\bf{g}}_u}{{\bf{\Psi }}_T}{{\bf{h}}_R}\sqrt {{L_{Riu}}} } \right|^2}\\
{\rm{subject}}\;{\rm{to}}\; & {\beta _{1,R}} \cdots {\beta _{N,R}} = 0.5\\
& {\beta _{1,T}} \cdots {\beta _{N,T}} = 0.5\\
& {\theta _{1,R}} \cdots {\theta _{N,R}} \in \left[ {0,2\pi } \right)\\
& {\theta _{1,T}} \cdots {\theta _{N,T}} \in \left[ {0,2\pi } \right).
\end{aligned}
\end{array}
\end{equation}

Therefore, by employing the signal alignment technique, as detailed in~\cite{t35}, our goal can be realized to effectively enhance the received power. To configure the phase shifts appropriately, the reflection channel vector and the transmission channel vector are defined as follows:
\begin{equation}\label{reflection channel matrix }
{{{\bf{\tilde h}}}_R} = \left[ {\begin{array}{*{20}{c}}
{{g_{R,1u}}{h_{R,i1}}}& \cdots &{{g_{R,Nu}}{h_{R,iN}}}
\end{array}} \right],
\end{equation}
and
\begin{equation}\label{transmission channel matrix }
{{{\bf{\tilde h}}}_T} = \left[ {\begin{array}{*{20}{c}}
{{g_{T,1u}}{h_{R,i1}}}& \cdots &{{g_{T,Nu}}{h_{R,iN}}}
\end{array}} \right].
\end{equation}

Thus, the reflection and transmission phase shifts arrays of the STAR-RISs can be further transformed into
\begin{equation}\label{reflection RIS phase shift design}
{{\bf{\Phi }}_R}{\rm{ = }}{\theta _{\rm{d}}} - \arg ({{{\bf{\tilde h}}}_R}),
\end{equation}
and
\begin{equation}\label{transmission RIS phase shift design}
{{\bf{\Phi }}_T}{\rm{ = }}{\theta _{\rm{d}}} - \arg ({{{\bf{\tilde h}}}_T}),
\end{equation}
where $\arg(\cdot)$ denotes the angle of the element, and ${\theta _{\rm{d}}}$ represents the desired phase, which can be set to any value. By implementing this approach, optimal channel gain is ensured for both indoor and outdoor users, allowing the objective function specified in \eqref{objective function CPA_reformulate} to be successfully achieved.

Then, the achievable ergodic rate of the outdoor and indoor users serving by the $i$-th INAC satellite can be given by:
\begin{equation}\label{ergodic rate_reflected user}
\begin{aligned}
&{R_{{\rm{R}},i}}  =\\
& {\log _2}\left( {1 + \frac{{{{\left( {\sum\limits_{n = 1}^N {} \left| {{\beta _{1,R}}{g_{R,1u}}{h_{R,i1}}\sqrt {{L_{Riu}}} } \right|  + \left| {{h_{iu}}\sqrt {{L_{iu}}} } \right|} \right)}^2}}}{{{\sigma ^2}}}} \right),
\end{aligned}
\end{equation}
and
\begin{equation}\label{ergodic rate indoor user}
{R_{{\rm{T}},i}} = {\log _2}\left( {1 + \frac{{{{\left( {\sum\limits_{n = 1}^N {} \left| {{\beta _{1,T}}{g_{T,1u}}{h_{R,i1}}\sqrt {{L_{Tiu}}} } \right|} \right)}^2}}}{{{\sigma ^2}}}} \right).
\end{equation}

We then can compare the communication performance of different satellites, where the satellite providing the maximum achievable ergodic rate can be selected for the CPA as shown in \textbf{Algorithm~\ref{Algorithm 2} }.

\begin{algorithm}[t]
\caption{CPA to solve problem~\eqref{objective function CPA_reformulate}} \label{Algorithm 2}
\begin{algorithmic}[1]
%
\STATE Establish ${{\bf{g}}_u}$, ${{\bf{\Psi }}_R}$, ${{\bf{h}}_R}$, ${{\bf{h}}_{iu}}$, $\sqrt {{L_{Riu}}}$ and $\sqrt {{L_{iu}}} $ for all available INAC satellites \\
\STATE Initialize ${{\bf{\Phi }}_R}$ and ${{\bf{\Phi }}_T}$\\
    \REPEAT
    \STATE{Obtain the reflection and transmission passive beamforming matrix of STAR-RIS}
    \STATE Calculate the achievable ergodic rate of user by each INAC satellite.
    \UNTIL{All the satellites have been traversed.}
    \STATE Select the INAC satellite which maximizes the achievable ergodic rate.
\end{algorithmic}
\end{algorithm}

\begin{remark}\label{remark3: CO-INAC}
The results in~\textbf{Algorithm~\ref{Algorithm 2}} indicate that by utilizing SIC technique, the CO-INAC scenario is capable of providing higher achievable ergodic rate in the high SNR regimes.
\end{remark}

\begin{remark}\label{remark4: Communication satellite expectation}
The results in~\textbf{Algorithm~\ref{Algorithm 2}} demonstrate that in order to improve the achievable ergodic rate performance, the nearest satellite is expected for the CPA.
\end{remark}

\section{Numerical Results}

In this section, numerical results are provided for the performance evaluation of the proposed NPA and CPA assisted INAC networks. As shown in Fig.~\ref{RISm}, the STAR-RIS is located at the window of the building to provide access services to indoor and outdoor users. The iteration number of the least squares method is set to 2000 times, where the accuracy is set to 0.1 meters. We then select 10 available satellites, where the locations are shown in Table~\ref{simulation parameter setting}.

\begin{table*}
\caption{\\ TABLE OF LOCATIONS}
\centering
\begin{tabular}{|l|l|l|l|}
\hline
Target & X-axis coordinates & Y-axis coordinates & Z-axis coordinates \\
\hline
STAR-RIS & 2451473.43334794 & 2940007.18127632 & 5084877.94326077 \\
\hline
Indoor user & 2451523.43334794 & 2940057.18127632 & 5084857.94326077  \\
\hline
Outdoor user & 2451423.43334794 & 2939957.18127632 & 5084827.94326077 \\
\hline
Satellite 1 & 2384140.77986545 & 26292387.6749704 & -1752765.80294385 \\
\hline
Satellite 2 & -7688937.22670325 & 13088957.6457098 & 21791665.4813813 \\
\hline
Satellite 3 & 7694983.70804847 & -12857727.5493792 & 22058611.9934355 \\
\hline
Satellite 4 & 21593131.9113028 & 14858836.7899355 & -4809198.45852993 \\
\hline
Satellite 5 & 14735759.3485476 & 3642752.94843750 & 21710269.2023414\\
\hline
Satellite 6 & 10822949.9268744 & 17448224.4300194 & 16861015.1148962 \\
\hline
Satellite 7 & 22983405.0752494 & -2550895.23789826 & 13042468.3643485 \\
\hline
Satellite 8 & 15960648.1354986 & -4443134.15738840 & 20811348.4358723 \\
\hline
Satellite 9 & 23113652.8643512 & 1123278.14965420 & 6871538.15438270 \\
\hline
Satellite 10 & 16937593.1345824 & -14466934.1345798 & -14539112.5683248 \\
\hline
\end{tabular}
\label{simulation parameter setting}
\end{table*}

The path loss exponents of  $\varsigma_i - STAR-RIS$, $\varsigma_i - user_u$, as well as $STAR_RIS - user_u$ links are set to 2, 2, as well as 2.2, respectively. The transmit power at each satellite is set to 40 W. The transmit antenna gain is set to 30 dB. The carrier frequency is set to 1 GHz. The bandwidth is set to $BW=10$ MHz. The power of AWGN is set to $-174+10{\rm{log}}(BW)$.
The reflection and transmission phases of STAR-RIS are continuous, and the amplitude coefficient of each RIS element is identity. For simplicity, we set the reflection amplitude coefficient and transmission amplitude coefficient to 0.5.
The small-scale fading parameters of the shadow Rician fading channels are set to $b = 0.279$, $m = 2$, as well as $\Omega = 0.251$.
In order to better illustrate the performance, we do not consider the effect of atmospheric time delay and other influencing factors.

\subsection{Impact of Satellite Selection for Positioning Accuracy}

In this subsection, we evaluate the positioning error of both the indoor and outdoor users by NPA in Fig.~\ref{figure_A} for 500 independent realizations of the instantaneous noise. The number of visible satellites of outdoor users and indoor users is set to 3 and 0, respectively. On the one hand, in Fig.~\ref{figure_A}, the STAR-RIS connects one satellite to the outdoor user for providing positioning services. On the other hand, since there is no direct link between satellites and indoor users, the STAR-RIS connects four satellites to the indoor user. Observe that when the PDoP increases, the positioning error of both indoor and outdoor users increases. This is due to the fact that high PDoP indicates poorer geometry distribution of satellites, resulting in higher positioning errors.
As can be seen from the red and purple curves in Fig.~\ref{figure_A}, the positioning error of indoor users is much higher than that of the outdoor user, which is because that the STAR-RIS to indoor user link is shared by all satellites. For example, when the distance between STAR-RIS and the indoor user is 10 meters, and compared to the positioning error of outdoor users, the positioning error of the indoor user also increases by 10 meters. This phenomenon also verifies our {\textbf{Remark~\ref{remark1: Indoor share path}}}.

\begin{figure}[ht]
\centering
\includegraphics[width =3in]{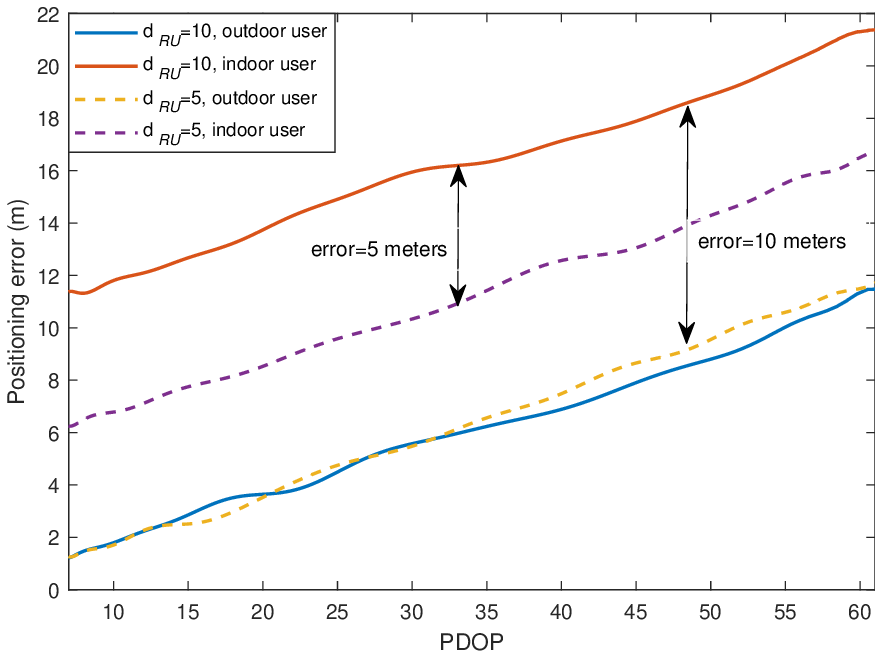}
\caption{The positioning error versus PDoP of both outdoor and indoor users. }
\label{figure_A}
\end{figure}

Here, we evaluate the positioning error of both the indoor and outdoor users with the aid of NPA with different number of satellites for 500 independent realizations in Fig.~\ref{figure_B}. For comparison, we denote the random satellite selection algorithm as ``RSA'' in Fig.~\ref{figure_B}, for which we assume that the STAR-RIS connects a random satellite.
In Fig.~\ref{figure_B}, there are a total of 11 available satellites in the proposed INAC networks. For the outdoor users, there are at least 3 visible satellites, whereas there is no visible satellite for the indoor users. As we can observe from Fig.~\ref{figure_B}, the proposed NPA has superior performance than that of the RSA for both indoor and outdoor users. We can also observe that indeed, more satellites connecting is capable of improving the positioning performance. However, the minimum positioning error of the indoor user is higher than 5 meters, which indicates that for the indoor users, no matter how many satellites are connected to the indoor users through STAR-RIS, the positioning error mainly depends on the distance between STAR-RIS and indoor users, which also verifies both {\textbf{Remark~\ref{remark1: Indoor share path}}} and {\textbf{Remark~\ref{remark2: navigation satellite expectation}}}.

\begin{figure}[ht]
\centering
\includegraphics[width =3in]{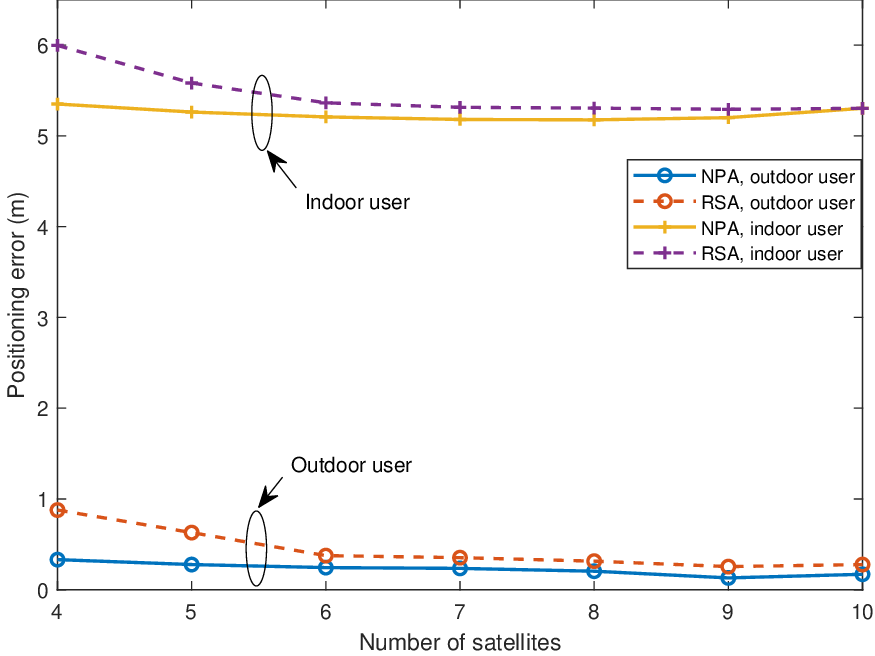}
\caption{The positioning error versus the number of satellites, where the distance between STAR-RIS and both indoor and outdoor user is set to 5 meters. }
\label{figure_B}
\end{figure}

\subsection{The Impact of Satellite Selection for Transmit Power}

\begin{figure}[ht]
\centering
\includegraphics[width =3in]{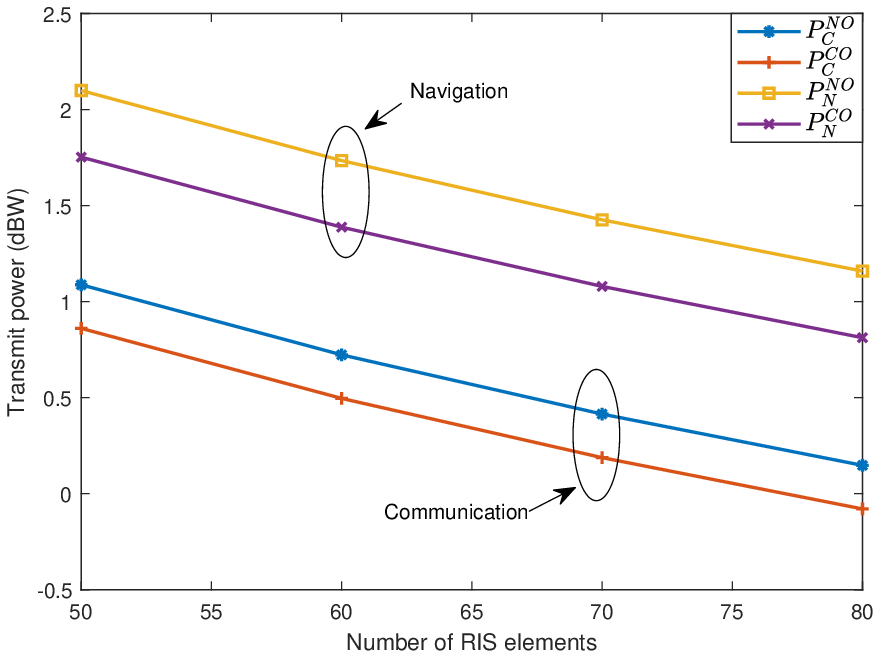}
\caption{The minimum transmit power versus the number of STAR-RIS elements. }
\label{figure_C}
\end{figure}

In Fig.~\ref{figure_C}, we focus our attention on the minimum transmit power with different numbers of STAR-RIS elements in both the NO-INAC and CO-INAC scenarios. The minimum rates of the indoor and outdoor users are set to $1$ and $2$ bits, respectively. The minimum required SNR threshold is set to $0.7$, which is 5.3 dB. The power allocation factors are respectively set to $\omega_C=0.65$ and $\omega_N=0.35$. As can be seen from Fig.~\ref{figure_C}, when more STAR-RIS elements are deployed, the required transmit power decreases. This is obviously due to the fact that the signal transmitted or reflected by STAR-RIS can be perfectly boosted at users, resulting in a lower transmit power requirement. On the other hand, we can also observe that in both CO-INAC and NO-INAC scenarios, the transmit power requirement of navigation signals is higher than that of the communication signals, which indicates that more power is expected for the navigation signal.

\subsection{The Impact of directly connected satellites on PDoP}
\begin{figure}[ht]
\centering
\includegraphics[width =3in]{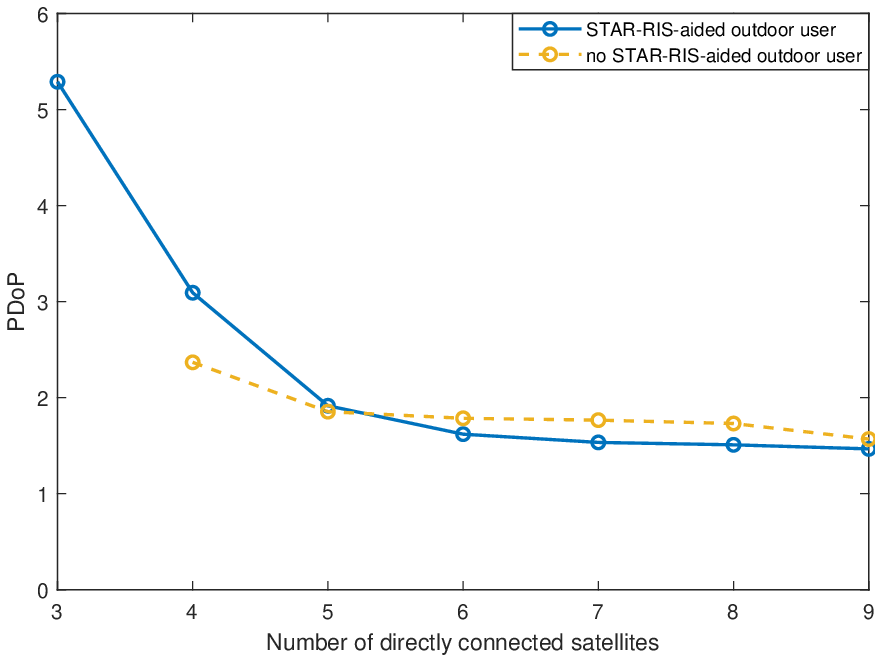}
\caption{The PDoP versus the number of directly connected satellites. }
\label{figure_D}
\end{figure}

Let us continue to analyze the number of directly connected satellites on PDoP in Fig.~\ref{figure_D}. In order to better illustrate the performance of the proposed networks, the PDoP of the outdoor user without STAR-RIS is also considered, denoted as ``no-RIS''. Note that when there is no STAR-RIS, and the number of directly connected satellites is 3, the PDoP does not exist. On the contrary, with the aid of STAR-RIS, only 3 directly connected satellites are capable of providing navigation services, which illustrates the superiority of the proposed satellite selection networks. By comparing two curves in Fig.~\ref{figure_D}, we can also observe that the PDoP floor of the STAR-RIS-aided networks is lower than that of ``no-RIS'' scenario, which is because that the location of STAR-RIS also improves the PDoP performance.

\subsection{The Achievable Ergodic Rate}

\begin{figure}[ht]
\centering
\includegraphics[width =3in]{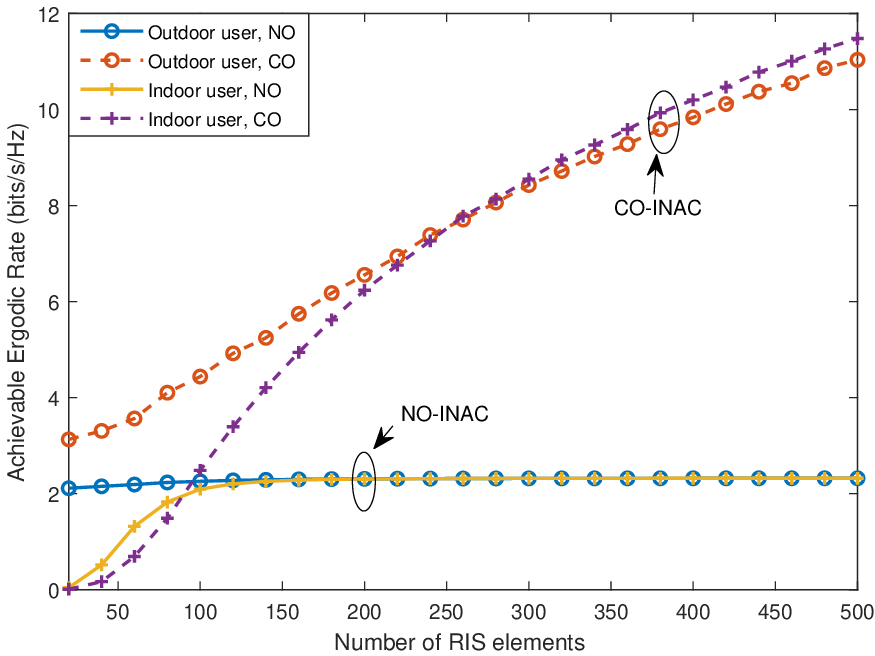}
\caption{The achievable ergodic rate of the indoor and outdoor users in both the CO-INAC and NO-INAC scenarios. }
\label{figure_EF}
\end{figure}

In Fig.~\ref{figure_EF}, we focus our attention on the achievable ergodic rate in the NO-INAC and CO-INAC scenarios. The power allocation factors are respectively set to $\omega_C=0.8$ and $\omega_N=0.2$. The achievable ergodic rates of the indoor and outdoor users are averaged by 1000 independent Monte Carlo simulations. Note that we only focus on the achievable ergodic rate of communications, where the navigation rate is ignored in Fig.~\ref{figure_EF}. Observe that in the NO-INAC scenarios, the achievable ergodic rate ceilings of both the indoor and outdoor users occur, which indicates that the NO-INAC scenario is only suitable for the case with low data rate requirements. On the contrary, we can see that the achievable ergodic rate of both the indoor and outdoor users increases with the aid of STAR-RIS arrays. In addition, the achievable ergodic rate of the indoor user is higher than that of the outdoor user when the number of STAR-RIS elements is high enough. This is because that the STAR-RIS is closer to the indoor user, leading to the higher SNR in the high STAR-RIS element case. The simulation results in Fig.~\ref{figure_EF} verify our {\textbf{Remark~\ref{remark3: CO-INAC}}}, where CO-INAC scenario is expected for higher achievable ergodic rate performance.

\begin{figure}[ht]
\centering
\includegraphics[width =3in]{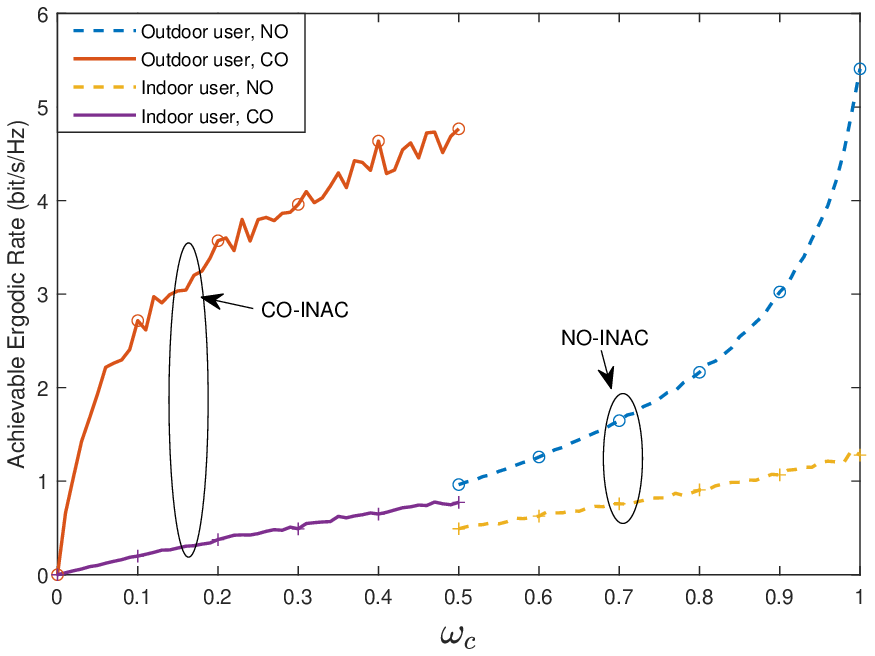}
\caption{The achievable ergodic rate versus the power allocation factors. }
\label{figure_G}
\end{figure}

We then turn our attention to the impact of power allocation factors. It is noted that in the CO-INAC scenarios, the power allocation factor of communication is lower than 0.5, whereas the power allocation factor of communication is higher than 0.5 in the NO-INAC scenarios. The transmit power is set to 46 dBm, and the number of STAR-RIS elements is set to 50. As we can see from Fig.~\ref{figure_G}, it is observed that the achievable ergodic rate of CO-INAC has a performance ceiling. On the contrary, the achievable ergodic rate increases with the increasing power allocation factor in the NO-INAC scenarios. This is due to the fact that the SIC technique has to first detect the signal with the higher power, and treat the signal with poorer power as interference. This phenomenon also indicates that a hybrid CO/NO-INAC selection algorithm is preferable. Moreover, since the power allocation factor cannot be higher than 1, the achievable ergodic rate performance of the CO-INAC scenarios outperforms than that of the NO-INAC scenarios, which confirms our {\textbf{Remark~\ref{remark3: CO-INAC}}}.

\begin{figure}[ht]
\centering
\includegraphics[width =3in]{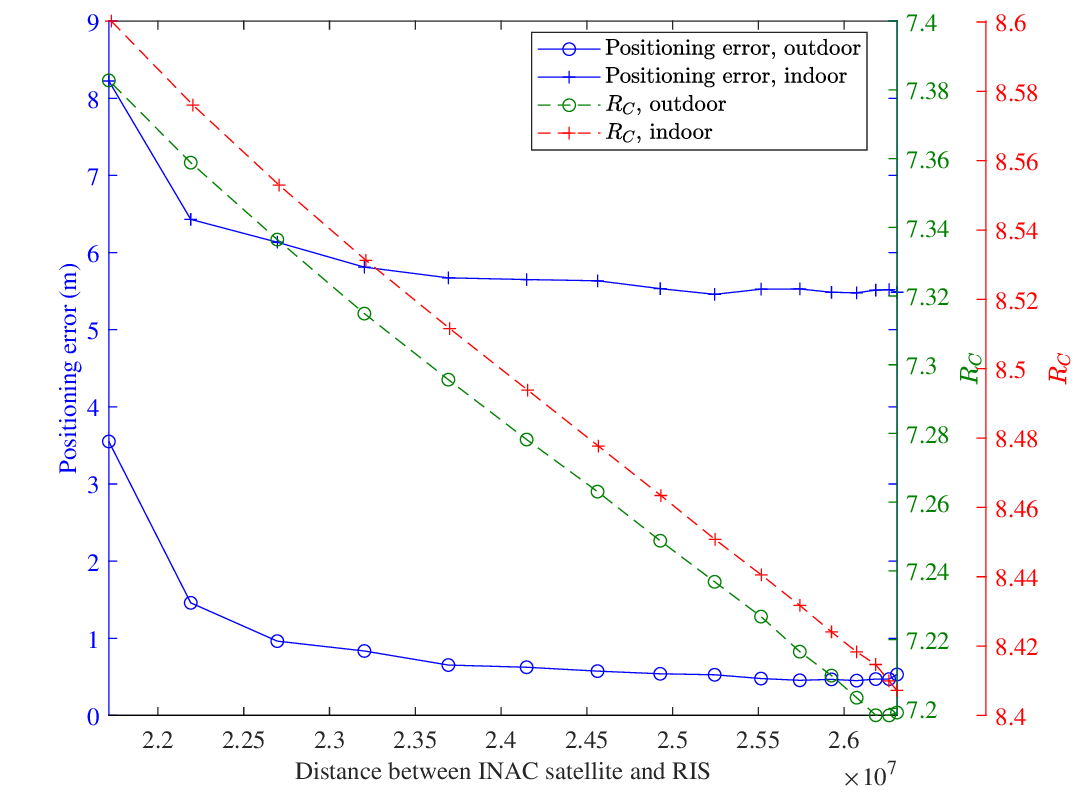}
\caption{The comparison between positioning error and achievable ergodic rate versus the distance between INAC satellite and STAR-RIS. }
\label{figure_H}
\end{figure}

In Fig~\ref{figure_H}, We then compare the impact of distance between the INAC satellite and STAR-RIS to the positioning error and achievable ergodic rate in the CO-INAC scenarios. Note that only one INAC satellite and three navigation satellites are available, where the height of the orbit are identical. Therefore, when the distance between the INAC satellite and STAR-RIS increases, where the elevation angle between the INAC satellite and ground user decreases, the PDoP can be decreased, resulting in a lower positioning error. On the contrary, we can observe that when the distance between the INAC satellite and RIS increases, the achievable ergodic rate decreases due to the increasing large-scale fading. In addition, with the increasing distance, the positioning error has its performance floor, whereas the achievable ergodic rate decreases with the increasing distance. This phenomenon also indicates the trade-off between positioning error and the achievable ergodic rate of the proposed satellite selection algorithm, which confirms both {\textbf{Remark~\ref{remark2: navigation satellite expectation}}} and {\textbf{Remark~\ref{remark4: Communication satellite expectation}}}.

\section{Conclusion}

In this article, we introduced a STAR-RIS-aided INAC network to address the challenges of urban canyon positioning. Initially, we outlined the existing difficulties faced by current positioning technologies and discussed the benefits of utilizing STAR-RIS. Building on traditional models, we developed a STAR-RIS-aided INAC network, strategically designing the NPA and CPA of the STAR-RIS array. We evaluated the network's performance by analyzing the positioning error and achievable ergodic rate, demonstrating its effectiveness in both navigation and communication. This investigation lays the groundwork for future advancements in STAR-RIS technology to meet the increasing demands for accurate and reliable GNSS services in obstructed environments. The findings indicate that the positioning error of the indoor user mainly depends on the distance between STAR-RIS and indoor users. A potential future direction is to minimize the positioning error of indoor users by merging more available information.


\bibliographystyle{IEEEtran}
\bibliography{IEEEabrv,bib2014}

\end{document}